\documentstyle[12pt]{article}
\begin{document}
\title{Scaling of Particle Trajectories on a Lattice II: The 
Critical Region}
\author{ Meng-She Cao and E.G.D. Cohen \\
The Rockefeller University \\
1230 York Avenue, New York, NY 10021} 
\maketitle
\vspace{0.3in}
\begin{center} \bf Abstract \end{center}

The scaling behavior of the closed trajectories of a moving 
particle generated by randomly placed rotators or mirrors on a square
 or triangular lattice in the critical region are investigated.
  We study numerically two scaling functions: $f(x)$ related to 
the trajectory length distribution $n_S$ and $h(x)$ related to 
the trajectory size $R_S$ (gyration radius) as introduced by 
Stauffer for the percolation problem, where $S$ is the length of a 
closed trajectory. The scaling function $f(x)$ is in most cases 
found to be symmetric double Gaussians with the same characteristic
size exponent $\sigma=0.43\approx 3/7$ as was found at criticality. 
In contrast to previous assumptions of an exponential 
dependence of $n_S$ on $S$, the Gaussian functions lead to a stretched
 exponential dependence of $n_S$ on $S$, $n_S\sim e^{-S^{6/7}}$. 
However, for the  rotator model on the partially occupied 
square lattice, an alternative scaling function near criticality is found, 
leading to a new exponent $\sigma '=1.6\pm0.3$ and a 
super exponential dependence of $n_S$ on $S$. The appearance of the same 
exponent $\sigma=3/7$ describing the behavior at and near the critical 
point is discussed. Our numerical simulations show that $h(x)$ is 
essentially a constant, which depends on the type of lattice and on 
the concentration of the scatterers. 

\vspace{1.5cm}

\newpage
\pagestyle{plain}
\baselineskip=1.5\baselineskip

\section{Introduction}
In a previous paper$^{(1)}$ we studied numerically the scaling 
behavior of extented (possibly infinite) particle trajectories 
generated by randomly distributed (rotator or mirror) scatterers 
on a square or a triangular lattice. We restricted ourselves there
 to the scaling behavior of structural properties of particle 
trajectories --- such as of the asymptotic number of right rotators
 on a particle trajectory or of the asymptotic number of sites
 visited by the moving particle on its trajectory with different
 frequencies --- that occurred strictly {\em at} criticality. In
 this paper we will discuss the scaling behavior found in the 
region {\em near} criticality, i.e. we study the critical behavior
 when one approaches the critical point from the outside. This 
involves, in addition to the scaling exponents $\tau$, $d_f$ and 
$\sigma$, which already appeared in previous papers$^{(1-4)}$, 
the determination of a scaling function defined by:
\begin{eqnarray}
n_{S} = S^{-\tau+1}f[ (C_{R}-C_{R_c}) S^{\sigma}] 
\label{ch1eq1}
\end{eqnarray} 
where $n_{S}$ is the probability to find a closed trajectory of 
length $S$, $C_{R}$ and $C_{R_{c}}$ are the concentration (i.e. 
the fraction) of right scatterers and the critical concentration
 of right scatterers on the lattice, respectively. For both the 
square and the triangular lattices fully occupied by rotators, 
one can show , by mapping the rotator model onto a percolation 
problem$^{(5,6)}$ that has been solved exactly before$^{(8)}$, that 
$C_{R_{c}}=1/2$, $\tau=15/7$, $d_f=7/4$ and $\sigma=3/7$. The 
scaling function $f(x)$ yields more detailed information about the
 trajectory size distribution than the exponent $\sigma$, where 
$x=(C_R-C_{R_c})S^{\sigma}$. Since there are no infinitely extended
trajectories away from criticality, $f(x)$ must vanish when 
$x\rightarrow \infty$. The exponent $-\tau+1$ instead of the usual
 $\tau$ occurs because the trajectories are constructed here 
from, what corresponds in percolation theory to, a seed and the 
number of choices to put the seed on a trajectory is proportional
 to the size of the trajectory itself$^{(1,2,9,10)}$. When $x=0$
 and $f(x)$ becomes a constant, we have, from Eq.~(\ref{ch1eq1})
, $n_S\sim S^{-\tau+1}$, recovering the results obtained previouly
 at criticality$^{(1,2)}$.      

One important feature about the scaling function $f(x)$ is that 
it is essentially invariant once we get into the critical region,
 i.e. the scaling function we obtain at one concentration near 
criticality is essentially the same as that obtained at another 
concentration. Note that since $x=(C_R-C_{R_C})S^{\sigma}$, to 
make $f(x)$ invariant, i.e. concentration independent, we need to
 give $C_{R_C}$ and $\sigma$ each a unique value. This provides 
an alternative method to compute both the exponent $\sigma$ and 
the critical concentration $C_{R_c}$, just from numerical data 
obtained at a few different concentrations in the critical region.
 
In addition to $f(x)$, there is another scaling function $h(x)$ 
introduced by Stauffer$^{(11)}$ some time ago,
\begin{eqnarray}
R_S=S^{1/d_f}h[ (C_{R}-C_{R_{c}}) S^{\sigma}] \label{hx}
\end{eqnarray}
Here $R_S$ is the gyration radius of a closed trajectory of length
$S$, in other words, $R_S^2$ is defined as the mean square 
displacement from the origin of all trajectories of length $S$. 
The scaling function $h(x)$ characterizes corrections to the fractal
 dimension as one moves away from criticality. From our numerical
simulations, we found, however, that $h(x)$ is essentially a
 constant for all $x$. This implies that $R_S \sim S^{2/d_f}$ 
holds quite  accurately in practice even for small trajectories, 
i.e. away from criticality. 

Previous numerical determinations of $\sigma$ were based on the 
calculation of the first moment of the trajectory length 
distribution$^{(2-4,12)}$, $< \! S \! >$, which is divergent at criticality,
\begin{eqnarray}
< \! S \!> &=&  \sum_{S}^{\infty} S n_S  \nonumber  \\
    &=&  \sum_{S}^{\infty} S^{-\tau+2} f((C_R-C_{R_c})S^{\sigma}
) \nonumber  \\
    &\sim& (C_R-C_{R_c})^{-(3-\tau)/\sigma} \label{avs}
\end{eqnarray}
Since $<\! S\! >\sim (C_R-C_{R_c})^{-\gamma}$ by definition, one
 has $\gamma=(3-\tau)/\sigma=2$. In this paper we compute the 
average of the mean square displacement of all trajectories, 
$< \! R^{2} \!>$, which also diverges at criticality,
\begin{eqnarray}
< \! R^2 \! > &=& \sum_{S}^{\infty}  R_S^2 n_S  \nonumber  \\
 &=& \sum_{S}^{\infty} S^{-\tau+1+2/d_f} f((C_R-C_{R_c})
S^{\sigma})h^2((C_R-C_{R_c})S^{\sigma})    \nonumber \\
&\sim& (C_R-C_{R_c})^{-(2+2/d_f-\tau)/\sigma}  \label{rsq}     
\end{eqnarray}
If we define an exponent $\rho$ by $<\! R^2 \! > \sim  (C_R-C_{R_c})^{-\rho}$
 one has $\rho=(2+2/d_f-\tau)/\sigma=7/3=2.333$. The hyperscaling
 relation in Eq.~(\ref{rsq}) is independent of the specific form
 of $f(x)$ and $h(x)$. Note that the upper limit of the summations
 in Eq.~(\ref{avs}) and  Eq.~(\ref{rsq}) is infinity, which 
poses a difficulty in getting very close to the critical point,  
since the dominant closed trajectories are then too large
 to generate on the computer. On the other hand, we can still 
compute the scaling function $f(x)$ by making a large cutoff
 ($2^{24}$) in $S$, so that trajectories whose length is bigger than 
$2^{24}$ are not considered. Although then the tail of $f(x)$ will
 get truncated by the cutoff, the remaining part still yields 
sufficient information to determine $\sigma$. 

The numerical algorithm used to generate the particle trajectories
 here has been explained in the previous paper$^{(1)}$. This a
lgorithm is very powerful due to its speed and its ability to   
generate large particle trajectories by using a virtual lattice 
scheme and a dynamic memory allocation technique. The calculation 
of each scaling function $f(x)$ or $h(x)$ and the calculation of 
$<\! S \!>$ or $<\! R^2 \!>$ at each concentration involve $50
0,000$ and $30,000$ particle trajectories, respectively.

The plan of this paper is the following. In section 2 the critical
 region for trajectories generated by rotators on a square 
lattice is discussed.  For the    critical behavior near $C_{R_{c}}
=C_{L_{c}}=1/2$, two qualitatively different scaling functions, 
$f(x)$ a double Gaussian and $f'(x)$ an exponential function, 
corresponding to two different values of $\sigma$: $\sigma
=0.43\approx 3/7$ and $\sigma'=1.6$, respectively, near this critical 
point have been determined, depending on how one approaches this point
 in the phase diagram. While $\sigma=0.43$ leads to a stretched
 exponential behavior of $n_{S}$ along the line $C_R+C_L=1$, 
$\sigma'=1.6$ leads to, what one could call, a super-exponential
 behavior along the line $C_R=C_L$. $h(x)$ is obtained by computing
 $f(x)h(x)$, which appears to be proportional to $f(x)$, so 
that $h(x)$ is essentially a constant. In section 3 we discuss the
 mirror model on the square lattice and a so-called quasi-rotator
 model deduced from the mirror model. In section 4, the critical
 region on the triangular lattice is discussed, which obtains 
both for rotators and mirrors. We found that the exponent 
$\sigma$ along the critical line is the same as that for $C=1$, i.e. 
$\sigma=0.43$, and that the scaling functions $f(x)$ and $h(x)$ 
are qualitatively the same and differ only by the values of the 
constants occurring in them. A discussion of the results obtained
 in this and the previous paper is given in section 5.

\section{Critical region for the rotator model on the square 
lattice}
\subsection{The fully occupied lattice}
As we have shown in the previous paper for $C=1$, the trajectories
 generated by a moving particle through randomly placed rotators
 can be mapped onto the perimeters of bond percolation clusters.
 Therefore the theory that has been developed for this percolation
 problem can be applied directly here. The critical concentration
 is $C_{R_{c}}=C_{L_{c}}=1/2$ and the exact values for 
$\sigma$, $d_f$ and $\tau$ are $3/7$, $7/4$ and $15/7$, respectively.
 From these known exponents and the critical concentration, we 
can compute the scaling function directly from Eq.~(\ref{ch1eq1}), 
\begin{eqnarray}
f((C_R-\frac{1}{2})S^{\sigma})&=& \frac{n_S}{S^{-\tau+1}}   \nonumber \\
		              &=& S^{8/7} n_S \label{ch2eq1}
\end{eqnarray}  
Note that since $n_S$ is the probability to find a trajectory of
 length $S$, the right hand side of Eq.~(\ref{ch2eq1}) is an average 
of $S^{8/7}$, taken over trajectories of length $S$ and can be 
easily determined in our numerical simulations. It is not difficult
 to see that $f(x)$ must be symmetric with respect to $x=0$, 
since the probability to generate the same trajectory is invariant
 under the transformation of interchanging $C_R$ and $C_L$, i.e.

\begin{eqnarray}  
f((C_R-\frac{1}{2})S^{\sigma})=f((C_L-\frac{1}{2})S^{\sigma})=f(
-(C_R-\frac{1}{2})S^{\sigma})
\end{eqnarray} 

Our numerical calculations of the scaling function $f(x)$ were 
carried out at $C_R=0.47$, $0.48$ and $0.49$, respectively. We 
found that the scaling functions obtained at these three concentrations
 collapse to a single curve, which could be fitted to a double
 Gaussian, i.e a sum of two overlapping Gaussians, (Fig. 1),
\begin{eqnarray}
f(x)=1.03 e^{-2.25(x+0.86)^2} +1.03 e^{-2.25(x-0.86)^2} 
\label{ch2eq2}
\end{eqnarray}  
Note that we determined $60$ values of $f(x)$ for each $C_R$, so
 that the curve in Fig. 1 contains $180$ points. When $x \gg 1$,
 $f(x)$ can be appoximated by
\begin{eqnarray}
f(x) \sim e^{-2.25 x^2}
\end{eqnarray} 
Therefore, since $n_S$ is proportional to $f(x)$, $n_S$ exhibits
 a stretched exponential decay for large $S$: $n_S\sim e^{-S^{6/
7}}$, in contrast to the exponential decay reported in the 
literature$^{(3,4)}$ based on the solution of the percolation 
problem on the Bethe lattice. 

In order to calculate $h(x)$, we first computed the product of 
$h(x)f(x)$. From Eq.~(\ref{ch1eq1}) and Eq.~(\ref{hx}), we have
\begin{eqnarray}
h(C_R-C_{R_c})f(C_R-C_{R_c}) = n_S \frac{R_S}{S^{\tau-1+d_f}} 
\label{ch2hx} 	
\end{eqnarray}
The right hand side of Eq.~(\ref{ch2hx}) is just the average of 
$R_S/S^{\tau-1+d_f}$ for fixed $S$, which can be easily calculated
 numerically. Our numerical results show then that $h(x)f(x)$ 
is proportional to $f(x)$, Eq.~(\ref{ch2eq2}), (Fig. 2), 
\begin{eqnarray}
f(x)h(x)=0.39 e^{-2.25(x+0.86)^2} +0.39 e^{-2.25(x-0.86)^2}
\label{ch2hx1}
\end{eqnarray}
so that $h(x)$ is essentially a constant, $0.37$.

We have also obtained the first moment of the trajectory length 
distribution as a function of $C_R-C_{R_c}$. Our numerical 
calculations show the following power law behavior,
\begin{eqnarray}
<S>= (C_R-C_{R_c})^{-\gamma}  \label{ch2eq3}
\end{eqnarray}
where $\gamma=2.00\pm 0.01$, (Fig.~3), in good agreement with the
 exact result $\gamma=(3-\tau)/\sigma=2$. Our numerical results
 for the mean square displacement of the trajectories also show 
a power law behavior,
\begin{eqnarray}
<R^2>\sim (C_R-C_{R_c})^{-\rho}  \label{ch2eq4}
\end{eqnarray} 
where $\rho=2.33\pm 0.01$, (Fig.~4), in good agreement with the 
exact result $\rho=(2+2/d_f-\tau)/\sigma=1/\sigma=7/3=2.333$.  

As a check, we also used a different method, the so called 
histogram method, to calculate the scaling function $f(x)$ for different
 concentrations from data computed at a given concentration 
$C_R$. A similar method has been used by Leath$^{(9,10)}$ to study
 the scaling behavior of percolation clusters and by Ferrenberg
 and Swendsen$^{(13)}$ to study the critical behavior of the 
Ising model. This calculation is based on the following observation.
 Each trajectory that is generated at one concentration of right
 rotators ($C_R$) and left rotators ($C_L$), can also be generated
 at other concentrations. However, the probability with which
 the trajectory is generated depends on the concentrations,
\begin{eqnarray}
P(N_R,N_L)=A \, C_R^{N_R} C_L^{N_L}  \label{ch2eq5}
\end{eqnarray}
where $A$ is the normalization factor for $P(N_R,N_L)$
and $N_R$ and $N_L$ are the number of right rotators and left 
rotators contained in the trajectory, respectively.  $A$ can also 
be defined as the total number of all different closed trajectories
 generated on the lattice, so that $A$ does not depend on the
 concentration. The probability to generate the same trajectory 
at another concentration $C'_R$ and $C'_L$ is
\begin{eqnarray}
P'(N_R,N_L) &=& A' \, {C'_R}^{N_R} {C'_L}^{N_L} \nonumber \\
 &=& P(N_R,N_L) \frac{A'}{A} (\frac{C'_R}{C_R})^{N_R} (\frac{C'_
L}{C_L})^{N_L}
\label{ch2eq6}
\end{eqnarray}
where $A'$ is the normalization factor for $P'(N_R,N_L)$. Note 
that if $C'_R/C_R>1$, then $C'_L/C_L<1$, or vice versa. To evaluate
 $(C'_R/C_L)^{N_R}$ and $(C'_L/C_L)^{N_L}$ separately for large
 $N_R$ and $N_L$ is not feasible on the computer, since one term
 is too large and the other term is too small. In our simulations,
 their product is computed from an alternative expression, 
$e^{N_R \, \ln(C'_R/C_R)+N_L \, \ln(C'_L/C_L)}$, which works quite
 well.  Here, we emphasize  that, although in theory $A$ and $A'$
 should be the same, for numerical simulations, $A$ and $A'$ are
 not equal due to systematic errors in the Monte Carlo samplings,
 so that they have be to be determined from Eq.~(\ref{ch2eq5})
 and Eq.~(\ref{ch2eq6}), respectively. $n_S$ at concentration 
$C'_R$ can be obtained from $P'(N_R,N_L)$ as, 
\begin{eqnarray}
n_S=\sum_{N_R} \sum_{N_L} P'(N_R,N_L) \delta(S-S'(N_R,N_L))  
\label{ch2eq7}
\end{eqnarray}
where $S'(N_R,N_L)$ is the length of a trajectory containing $N_
R$ left rotators and $N_L$ left rotators. Once $n_S$ is obtained,
 $f(x)$ can be calculated for any other concentration according
 Eq.~(\ref{ch2eq1}).

The standard data that we used were computed at the concentration
 $C_R=0.51$. From these we generated the scaling functions $f(x)$ 
for $C'_R=0.52$, $0.53$, $0.54$, $0.55$, all of which fall on
 the same curve (Fig. 5). Thus this method is consistent with the
 first one. These results also show that the critical region in
 concentration is at least as large as $0.05$.

\subsection{The partially occupied lattice}
For $C<1$, the particle trajectories cannot be mapped onto a 
percolation problem anymore, because the trajectories can cross 
themselves. It was found by Cohen and Wang$^{(5)}$ that there are 
two nonlinear critical lines symmetric with respect to the line $
C_R=C_L$,  merging at $C_R=C_L=1/2$, (Fig. 6). Moreover, we found
 that they appear to be tangent to the line $C=1$. Previous 
numerical results have shown that the trajectory fractal dimension 
$d_f$ and the exponent $\tau$ along the critical lines are the 
same as that in the case $C_R=C_L=1/2$. Here we show numerically,
 however, that in addition to the exponent $\sigma$ which we have
 investigated in the previous section, a different exponent 
$\sigma'$ also appears when $C<1$, suggesting a new scaling behavior.
   
For $C<1$, the simplest way to study the critical behavior near 
criticality is to approach the critical point $C_{R_c}=C_{L_c}=1
/2$ along the line $C_R=C_L$, which is perpendicular to the 
critical line at $C_R=C_L=1/2$. We assume that the cluster size 
distribution has a form similar to Eq.~(\ref{ch1eq1}),
\begin{eqnarray}
n_S=S^{-\tau+1}f'((1-C)S^{\sigma'})
\end{eqnarray}
where $f'(x)$ is a new scaling function in contrast to $f(x)$, 
which we have obtained before and $x$ is defined as $(1-C)
S^{\sigma'}$. Since $x$ cannot be negative, $f'(x)$ must be asymmetric 
with respect to $x=0$, in contrast to $f(x)$ of 
Eq.~(\ref{ch2eq2}). The exponent $\tau$, however, must be the same as before, 
since for $C=1$, the previously established relation $n_S\sim S^{
-\tau+1}$ must be recovered.  

The computation of the new scaling function $f'(x)$ is much more
 time consuming than that for the rotator model on the fully 
occupied lattice, although the procedure is similar. The reason is 
the following. When $C$ is small, the moving particle becomes less
 frequently scattered, therefore, on average, it takes longer 
for the particle trajectory to get closed. On the other hand, if
 $C$ is close to one, i.e. near criticality, the particle 
trajectories also close very slowly. On the line $C_R=C_L$, we found 
that the trajectories close fastest at $C_R=C_L=0.45$, when 
essentially, all trajectories are closed after $2^{21}$ time steps 
(Fig. 7). However, this concentration is not useful to calculate 
the scaling function, because it is not in the critical region.

As we explained in the introduction, if the critical behavior 
appears only for very large trajectories, direct calculation of 
$\gamma'$ and $\rho'$ --- the exponents corresponding to $\gamma$ 
and $\rho$, respectively --- from equations Eq.~(\ref{avs}) and 
Eq.~(\ref{rsq}) becomes very difficult, because the large 
trajectories are too large to generate on the computer. On the other 
hand, we can still compute the scaling function $f'(x)$ by making
 a cutoff, i.e. trajectories whose length is bigger than a fixed
 number are disregarded. In our simulations, we took this fixed 
number or cutoff as sixteen million time steps. By choosing 
$\sigma'=1.6$, we find that the scaling functions obtained for 
different concentrations, $C=0.99$, $0.985$, $0.98$,  all collapse to
 a single curve, (Fig. 8),
\begin{eqnarray}
f'(x)=0.475 e^{-(1.65\times 10^{-8} x)}
\end{eqnarray} 
where for each $C$, 25 points have been computed, so that the curve
 in Fig.~8 contains 75 points. Note that since $x\sim S^{1.6}$,
 $f'(x)$ decays as a super-exponential function of $S$, 
$n_S\sim e^{-S^{1.6}}$. In fact, eq.(17) obtains for values of $\sigma' =
1.6 \pm 0.3$. 

The other scaling function $h'(x)$ is obtained from the calculation
of $f'(x)h'(x)$, which we found again to be proportional to 
$f'(x)$, (Fig. 9),
\begin{eqnarray}
f'(x)h'(x)=0.115 e^{-(1.65\times 10^{-8} x)}
\end{eqnarray} 
so that $h'(x)=0.115/0.475=0.24$ independent of $x$. The exponents
 $\gamma'$ and $\rho'$ can be obtained from $\gamma'=(3-\tau)/
\sigma'=0.54$ and $\rho'=(2+2/d_f-\tau)/\sigma'=0.63$, respectively,
 both much smaller than those for $C=1$.
\section{Critical region for the mirror model on the square lattice}

It has been shown before, by mapping the mirror model to a 
percolation problem, that $C=1$ is a critical line$^{(5,6)}$, i.e. 
the size distribution and the fractal dimension of the trajectories
 have the same power law behavior as for the corresponding 
percolation problem at criticality, $n_S\sim S^{-\tau+1}\sim S^{-8/7}$
 or $\tau=15/7$ and $d_f=7/4$, respectively. However, the 
trajectory size average $<\! S \!>$ and the mean square displacement
 $<\! R^2 \!>$ are both divergent along the critical line. Therefore,
 the exponent $\sigma$ does not appear here. Although the 
mirror model can be mapped onto a quasi-rotator model which exhibits
 the same scaling behavior as that for the ordinary rotator 
model at criticality$^{(1)}$, the quasi-rotator model cannot be 
used to calculate $\sigma$, $\gamma$ and $\rho$, because at present
 we do not know how to generate the particle trajectories directly
 rather than deducing them from the mirror model. 

For $C<1$, the mirror model cannot be mapped onto a percolation 
problem anymore. It was found numerically that the critical 
exponents are drastically changed from those at $C=1$; $n_S\sim (\ln
 S)^{-1}$ and $d_f=2$ with logarithmic corrections$^{(5,6)}$. This
 new critical behavior exists in the whole $(C_R,C_L)$ plane, 
where super-diffusion occurs$^{(5)}$, except at the three boundary
 lines $C=1$, $C_R=0$ and $C_L=0$, where the moving particle 
simply zig-zags to infinity. Since $<\! S \!>$ and $<\! R^2 \! >$
 are both divergent everywhere, $\sigma$ cannot be defined either.
  
\section{Critical region for the rotator and the mirror model on
 the triangular lattice}
\subsection{The fully occupied lattice}
Since the rotator model and mirror model are equivalent on the 
triangular lattice$^{(1,5,7)}$, we only study the rotator model 
here. Similar to the rotator model on the square lattice, the 
rotator model on the triangular lattice can be mapped onto a 
percolation problem$^{(1,5,7)}$, a site percolation problem here, for 
which $\tau=15/7$, $d_f=7/4$ and $\sigma=3/7$, while the critical
 concentration is $C_{R_c}=C_{L_c}=1/2$. 

Our numerical calculations of the scaling function $f(x)$ were 
carried out at $C_R=0.47$, $0.48$ and $0.49$, respectively. The 
scaling functions for these three concentrations collapsed to a 
single curve, which could be fitted again to a double Gaussian, 
(Fig. 10),
\begin{eqnarray}
f(x)=0.94 e^{-1.80(x+0.96)^2} +0.94 e^{-1.80(x-0.96)^2} 
\label{ch5eq1}
\end{eqnarray}
where for each $C_R$, more than $60$ points were computed and the
 curve comprises therefore more than $180$ points. As before, 
$f(x)$ must be symmetric with respect to $x=0$, the reason being 
that the probability to generate the same trajectory is invariant
 under the transformation of interchanging $C_R$ and $C_L$. 
Eq.(7) and Eq.(19) are very similar escept 
that the constants are different.  When $x \gg 1$, $f(x)\sim e^{-
x^2}$, so $n_S\sim e^{-S^{6/7}}$ which is the same for the square
 lattice, i.e. the number of trajectories decays with a stretched
 exponential law in $S$.

The calculation of the scaling function $h(x)$ is similar to that
 for the square lattice. We found numerically, $f(x)h(x)$ is 
proportional to $f(x)$, (Fig. 11),
\begin{eqnarray}
f(x)h(x)=0.28 e^{-1.80(x+0.96)^2} +0.28 e^{-1.80(x-0.96)^2} 
\label{ch5eq2}
\end{eqnarray}
so that $h(x)$ is a constant, $0.28/0.94=0.30$.

The average trajectory size diverges as one approaches criticality 
with an exponent $\gamma$, $< \! S \!>\sim(C_R-C_{R_c})^{-\gamma}$.
 Our numerical simulations show that $\gamma=2.0\pm 0.01$, (Fig. 12), 
in good agreement with the exact result, $\gamma=2$. The
 mean square displacement of the particle trajectories diverges
 with an exponent $\rho$, $< \! R^2 \!>\sim(C_R-C_{R_c})^{-\rho}$,
 the value of $\rho$ obtained from our numerical simulations 
is $2.33\pm 0.01$ (Fig. 13), also in good agreement with the 
exact result $\gamma=7/3$.

Using the histogram method, we computed the scaling functions 
also from the standard data obtained at the concentration $C_R=0.51$,
 for the other concentrations $C_R=0.52$, $0.53$, $0.54$, $0.
55$. All the scaling functions fall on the same curve, 
Eq.~(\ref{ch5eq1}), (Fig. 14), showing that this method is consistent with
 the first one.
 
\subsection{The partially occupied lattice}
For $C<1$, the rotator model and the mirror model can be still 
mapped onto each other. However neither model can be mapped onto 
a percolation problem now. Here we only consider the rotator model.
 It was found before that there is only one critical line $C_
R=C_L$ rather than two critical lines as in the case of the rotator
 model on the  partially occupied square lattice$^{(1,5)}$.  
The critical exponents $\tau$ and $d_f$ along the critical line 
are found numerically to be the same as those at the critical point,
 $C_R=C_L=1/2$, viz. $15/7$ and $7/4$, respectively$^{(1,5,7
)}$. 

To study the critical behavior near criticality, we first need 
to choose the direction in which we approach criticality. The most
 obvious choice is the direction perpendicular to the critical 
line. Although the rotator model cannot be mapped onto a percolation
 problem, we found that, if we choose $\sigma=3/7$, the scaling
 functions computed at different $C_R$ collapse very well onto a single
 curve. This suggests that randomly distributed empty 
sites  on the lattice are irrelevant for the critical behavior. 
 The scaling functions were computed in the critical region near
 $C_{R_c}=C_{L_c}=0.425$ at three different $C_R$, $C_R=C_{R_c}+
0.01$, $C_R=C_{R_c}+0.015$, $C_R=C_{R_c}+0.02$ along the line $C
=0.85$.  We found that these scaling functions could be fitted 
to a double Gaussian, (Fig. 15),
\begin{eqnarray}
f(x)= 1.05 e^{-1.18(x+1.19)^2}+1.05 e^{-1.18(x-1.19)^2} 
\label{dtrifx}
\end{eqnarray}
where we computed $80$ points for each $C_R$, leading to $240$ 
points on the curve. From Eq.~(\ref{ch5eq1})and Eq.~(\ref{dtrifx}),
 one can see that all three constants contained in $f(x)$ are 
concentration dependent. 

The product of $f(x)h(x)$ is again found to be proportional to 
$f(x)$, (Fig. 16), 
\begin{eqnarray}
f(x)h(x)= 0.30 e^{-1.18(x+1.19)^2}+ 0.30 e^{-1.18(x-1.19)^2}
\end{eqnarray}
so that $h(x)=0.30/1.05=0.28$.

To investigate whether the critical exponent $\sigma$ has the same
 value along the critical line, we also computed $<\! S \!>$ 
and $< \! R^2 \!>$ in the critical region around $C_{R_c}=C_{L_c}
=0.45$ and $C_{R_c}=C_{L_c}=0.4$, respectively. In both cases, we
 found, from our numerical simulations, that the exponents 
$\gamma=2.00\pm 0.02$ and $\rho=2.33\pm 0.02$ are very close to their
 values at $C_{R_c}=C_{L_c}=0.5$: $2$ and $7/3$, respectively, 
(Fig. 17-18), suggesting that the critical behavior along the 
critical line $C_R=C_L$ belongs indeed to the same universality 
class.

\section{Conclusion}
We end with a few remarks.

\noindent 1. We have studied the critical behavior of particle 
trajectories on both the square lattice and the triangular lattice
 in the critical region, for both $C=1$ and $C<1$. Our methods 
were based on calculations of two scaling functions $f(x)$ (which
 yields $\sigma$) and $h(x)$, as well as the two exponents 
$\gamma$ and $\rho$ from $<\! S \! >$ and $<\! R^2 \! >$ respectively.
 From these two scaling functions, one derived and verified 
numerically the two scaling relations, $\gamma=(3-\tau)/\sigma$ 
and $\rho=(2+2/d_f-\tau)/\sigma$. Our results are summarized in 
table I.

\noindent 2. For $C=1$, the rotator model on the square lattice,
 and the rotator and the mirror model on the triangular lattice 
can all be mapped onto a percolation problem.  Our numerical 
results show that near $C_{R_c}=C_{L_C}=1/2$, $f(x)$ can be fitted
 to a sum of two overlapping Gaussians, i.e. a double Gaussian. 
Our numerical calculations of $<\! S \!>$ and $<\! R^2 \!>$ show
 that $\gamma=2.00\pm0.01$ and $\rho=2.33\pm 0.01$, respectively
, in excellent agreement with the exact results. Weinrib and 
Trugman$^{(14)}$ have argued that the scaling function $f(x)$ for 
the perimeter of percolation clusters should be proportional to 
the scaling function $g(x)$ for the percolation clusters themselves.
 It would be interesting to check this for the $f(x)$ we 
obained here.

\noindent 3. For $C<1$, however, the critical behavior of the 
rotator model on the square lattice and that of the rotator and 
mirror model on the triangular lattice are very different. None of
 these models can be mapped onto a percolation problem. For the 
square lattice, we obtained numerically a new exponent $\sigma'=
1.6\pm0.3$ which is significantly larger than the usual $\sigma=
3/7$. This indicates that the typical size of closed trajectories
 upon approach to criticality increases much slower to infinity
 along the line $C_R=C_L$ than along the line $C_R+C_L=1$. The 
new scaling function $f'(x)$, corresponding to the exponent 
$\sigma'=1.6$, appears to be an exponential function, rather than a  
double Gaussian as for $C=1$, but the origin of this exceptional
 critical behavior is unclear to us. For the triangular lattice,
 however, we found numerically, that $\sigma$ is very close to 
$3/7$ which also obtains for $C=1$, and that the scaling function
 $f(x)$ can still be described by a double Gaussian. This suggests
 that the critical behavior in the direction perpendicular to 
the critical line $C_R=C_L$ is in the same universality class. For
 dilute thermal systems, it is also found that the critical 
behavior is independent of the density of impurities if the specific
 heat exponent $\alpha$ is negative$^{(15)}$, but to what extent
 this is similar to the independence of the number of empty sites
 found here, is unclear to us. 

\noindent 4. The scaling function $h(x)$ appears to behave like 
a constant in all cases, indicating that there are no noticeable
 corrections to the universal fractal dimension $d_f=7/4$. This 
simply means that the gyration radius of the extended trajectories
 is independent of the degree of occupation of the lattice by scatterers
and only trivially dependent on the nature of the lattice.  

\noindent 5. We have studied in the previous paper the scaling 
behavior of extended particle trajectories as their sizes approach
 infinity {\em at} criticality. For example, we have obtained 
for the rotator model on the fully occupied square lattice that 
$<\! S/N-3/2 \!> \sim N^{-0.57}$, where the exponent is very close
 to $1-\sigma=4/7=0.571$, with $\sigma$ characterizing the 
trajectory size distribution in the critical region {\em near} 
criticality. We surmise that this may not be a coincidence and 
conjecture that there is a possible relation between the scaling 
behavior of extended closed trajectories {\em at} criticality and in 
the critical region {\em near} criticality. If so, the question 
arises whether also other exponents such as $\gamma$ and $\rho$ 
could be determined from the scaling behavior at criticality, rather
 than as we have done here, in the critical region. 

\noindent 6. Finally, we remark that the diffusive behavior of 
the moving particle away from criticality, which has been studied
 extensively before$^{(5,7)}$, is in fact controlled by the scaling
 function $f(x)$. The time dependent diffusion constant $D(t)$
 is defined as $D(t)=<\! R^2(t)\! >/(4t)= \sum_{}^{} R_S^2 n_S/
4t$, where the summation is taken over all closed trajectories upto
 length $t$.  For $(C_R-C_{R_c}) t^{\sigma}\ll 1$, when $f(x)
$ is essentially a constant, $D(t)$ is then virtually the same 
as at criticality, while for $(C_R-C_{R_c}) t^{\sigma}\gg 1$, 
when $f(x)$ is very small, $D(t)$ decays to zero as $1/t$. \\

\noindent{\bf Acknowledgement}\\
We acknowledge useful discussions with Prof. B. Nienhuis and 
finantial support under grant DE-FG02-88ER13847 of the Department 
of Energy.

\newpage
\noindent{\bf References}

\noindent 1. Meng-She Cao and E. G. D. Cohen, (previous paper).

\noindent 2. R. M. Ziff, Phys. Rev. Lett. {\bf{56}}, 545 (1986).

\noindent 3. M. Ortun\~{o}, J. Ruiz, and J. M. F. Gunn, J. Stat.
 Phys. {\bf{65}}, 453 (1991).

\noindent 4. J. D. Catal\'{a}, J. Ruiz, and M. Ortun\~{o}, Z. Phys.
 B {\bf{90}}, 369 (1993). 

\noindent 5. E. G. D. Cohen and F. Wang, J. Stat. Phys. {\bf{81}}, 445 
(1995).

\noindent 6. R. M. Ziff, X. P. Kong, and E. G. D. Cohen, Phys. Rev.
 B {\bf{44}}, 2410 (1991).

\noindent 7. X. P. Kong and E. G. D. Cohen, J. Stat. Phys. {\bf{62}}, 7
37 (1991).

\noindent 8. H. Saleur and B. Duplantier, Phys. Rev. Lett. {\bf{58}}, 2
325 (1987).

\noindent 9. P. L. Leath, Phys. Rev. Lett. {\bf{36}}, 921 (1976).

\noindent 10. P. L. Leath, Phys. Rev. B {\bf{14}}, 5046 (1976).

\noindent 11. D. Stauffer and A. Aharony, {\em Introduction to 
Percolation Theory}, London, Taylor and Francis (1992).

\noindent 12. R. M. Ziff, P. T. Cummings, and G. Stell, J. Phys.
 A {\bf{17}}, 3009 (1984).

\noindent 13. A. M. Ferrenberg and R. H. Swendsen, Phys. Rev. Lett.
 {\bf{61}}, 2635 (1988).

\noindent 14. A. Weinrib and S. Trugman, Phys. Rev. B {\bf{31}}, 2993 
(1985).

\noindent 15. A. B. Harris, J. Phys. C {\bf{8}}, 1671 (1974).

\newpage
\noindent{\bf Figure captions} \\

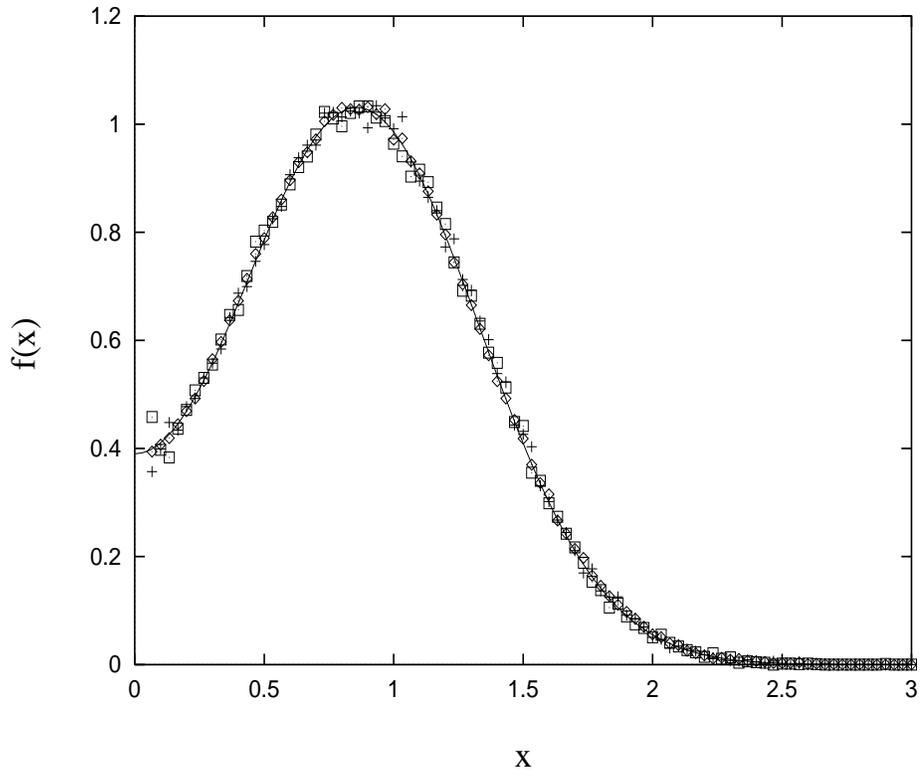
\noindent Fig. 1. The scaling function $f(x)$ vs $x$ for the rotator
 model on a fully occupied square lattice, computed at $C=C_
{R_c}+0.01$ ($\diamond$),  $C=C_{R_c}+0.02$ ($+$) and $C=C_{R_c}
+0.03$ ($\Box$). The curve is described by a double Gaussian, 
$f(x)=1.03 e^{-2.25(x+0.86)^2} +1.03 e^{-2.25(x-0.86)^2}$. The 
deviations between $f(x)$ and the numerical data near $x=0$ in this
 and later similar figures are due to the failure of scaling for
 small trajectories.
  
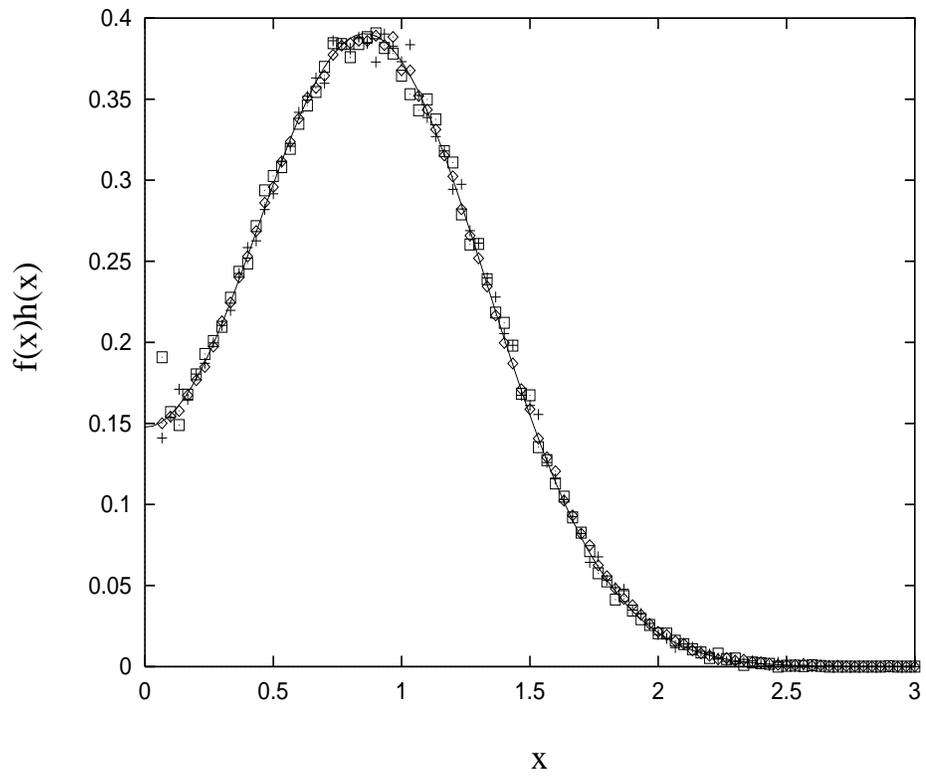
\noindent Fig. 2. The scaling function $f(x)h(x)$ vs $x$ for the
 rotator model on a fully occupied square lattice, computed at $
C=C_{R_c}+0.01$  ($\diamond$), $C=C_{R_c}+0.02$ ($+$) and $C=C_{
R_c}+0.03$ ($\Box$). The curve is described by a double Gaussian
, $f(x)h(x)=0.39 e^{-2.25(x+0.86)^2} +0.39 e^{-2.25(x-0.86)^2}$.

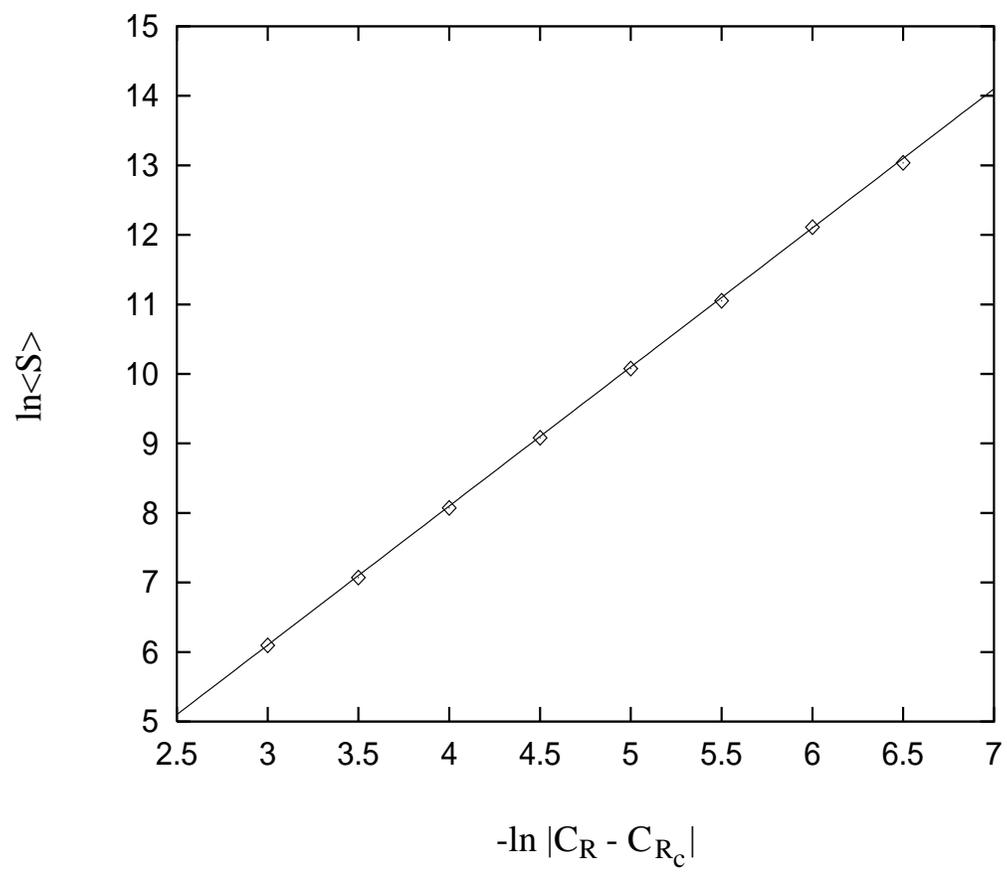
\noindent Fig. 3. $\ln<\! S \!>$ vs $-\ln(C_R-C_{R_c})$ for the 
rotator model on the fully occupied square lattice. The slope of
 the fitting line is $2.00$.

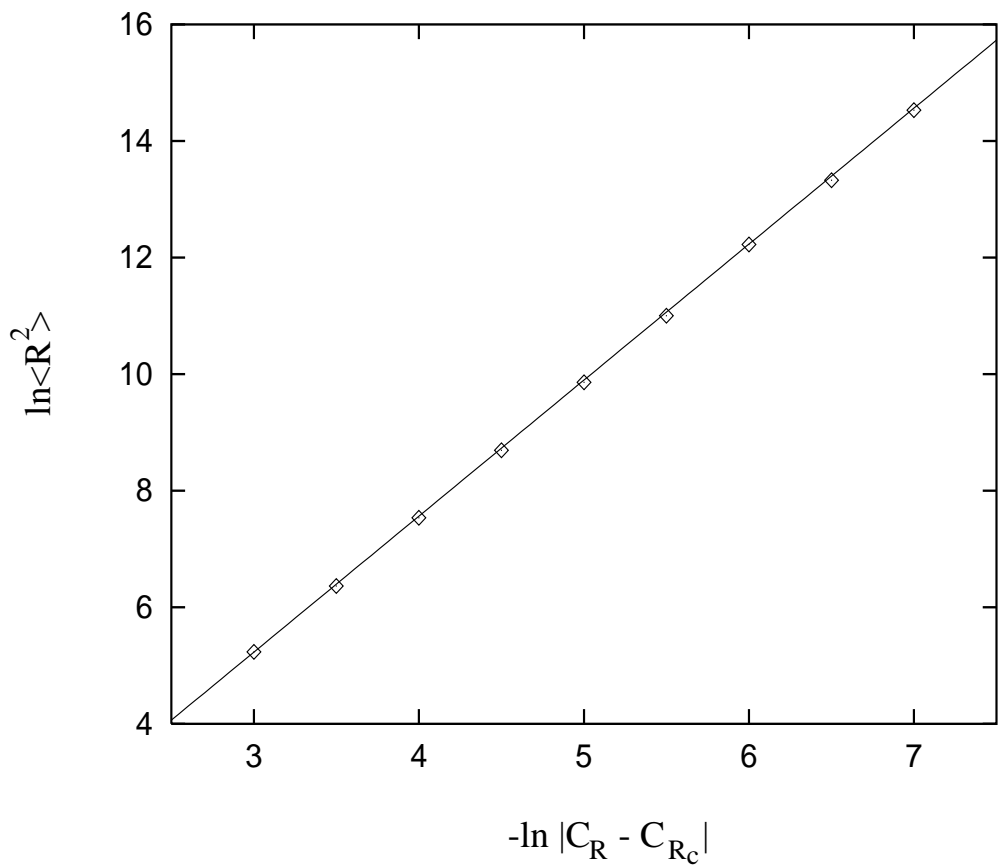
\noindent Fig. 4. $\ln<\! R^2 \!>$ vs $-\ln(C_R-C_{R_c})$ for the
 rotator model on the fully occupied square lattice. The slope 
of the fitting line is $2.33$.

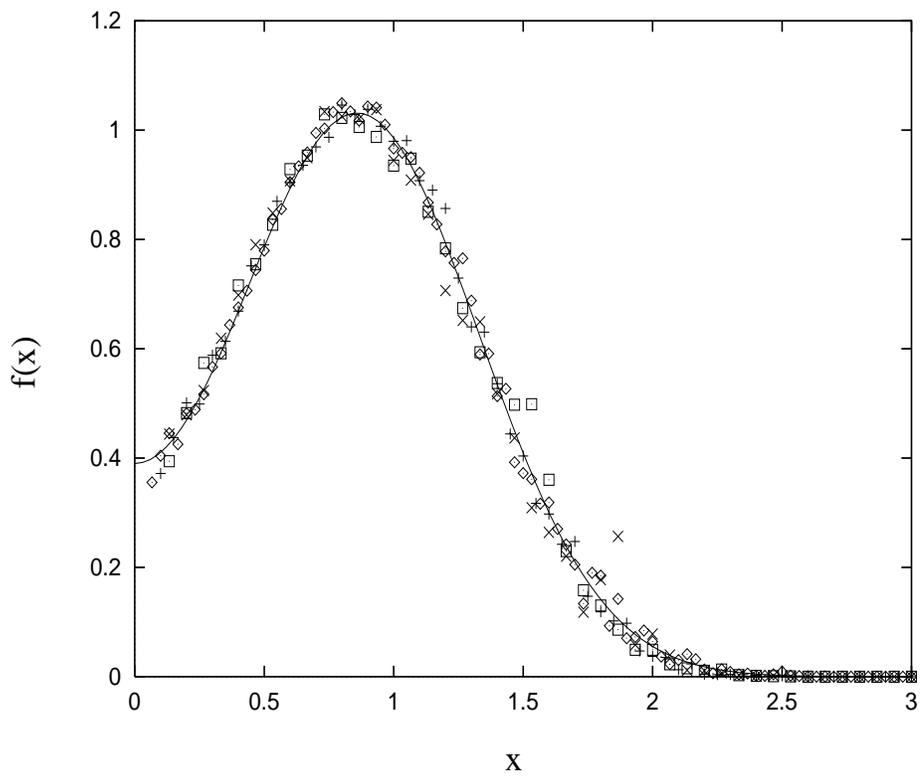
\noindent Fig. 5. The scaling function $f(x)$ vs $x$ for the 
rotator model on the fully occupied square lattice, obtained by 
using the histogram method. Computed for $C=C_{R_c}+0.02$ 
($\diamond$), $C=C_{R_c}+0.03$ ($+$), $C=C_{R_c}+0.04$ ($\Box$) and $C=C_
{R_c}+0.05$ ($\times$). The curve is described by a double Gaussian,
 $f(x)=1.03 e^{-2.25(x+0.86)^2} +1.03 e^{-2.25(x-0.86)^2}$.

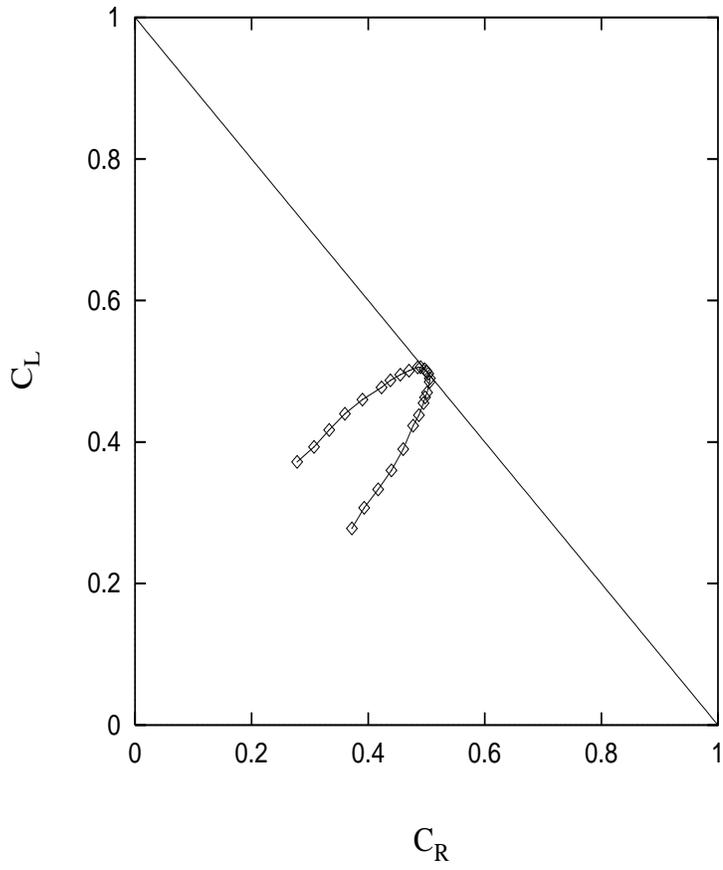
\noindent Fig. 6. Part of the phase diagram for the rotator model
 on the square lattice obtained from numerical simulations for 
$0.65\leq C\leq 1$. The lines are drawn to guide the eye.

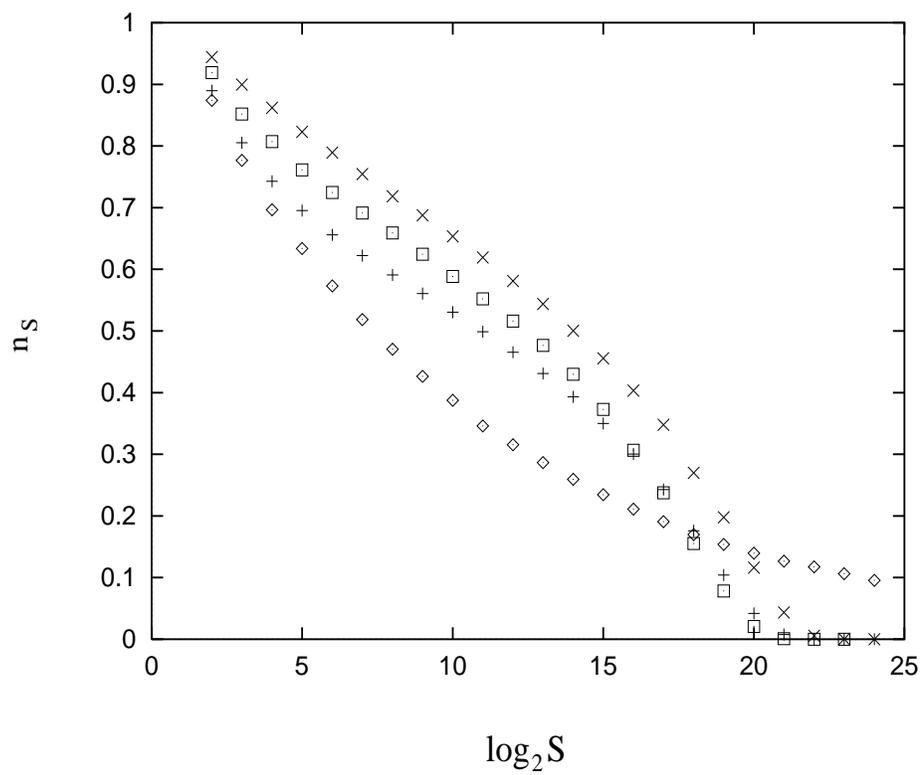
\noindent Fig. 7. $n_S$ vs $S$, at concentrations $C_R=C_L=0.5$ 
($\diamond$),  $C_R=C_L=0.485$ ($+$), $C_R=C_L=0.45$ ($\Box$) and
 $C_R=C_L=0.40$ ($\times$).

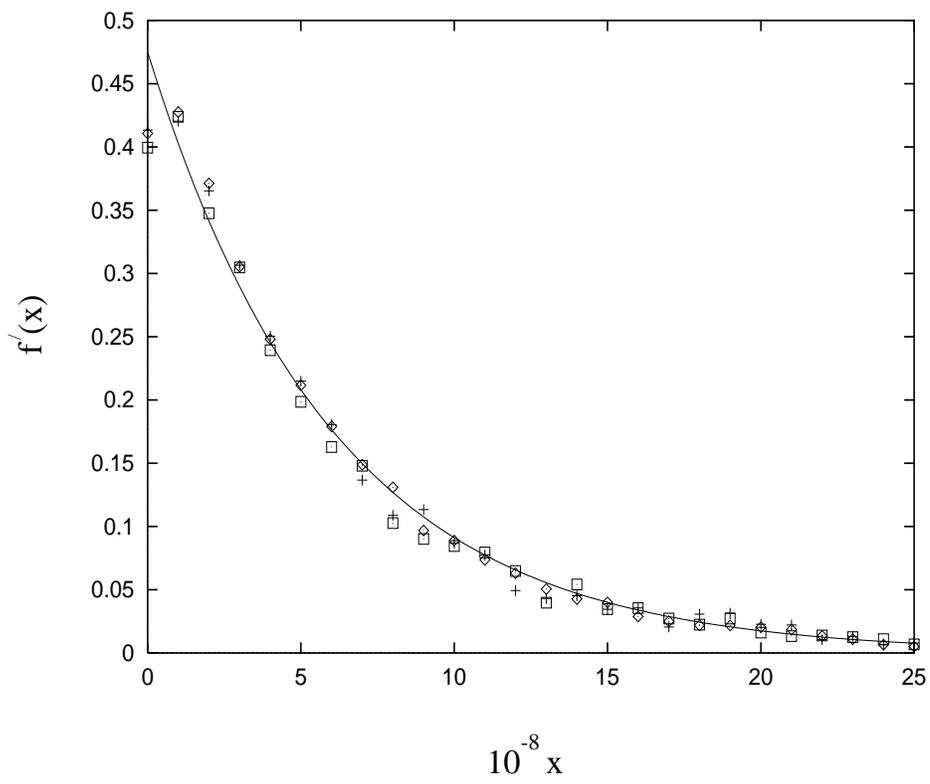
\noindent Fig 8.  The scaling function $f'(x)$ vs $x$ for the 
rotator model on the partially occupied square lattice, computed at
 $C=C_{R_c}+0.02$ ($\diamond$), $C=C_{R_c}+0.025$ ($+$) and $C=C
_{R_c}+0.01$ ($\Box$). The curve is described by an exponential 
function, $f'(x)=0.475 e^{-0.165 \times 10^{-8}x}$.
 
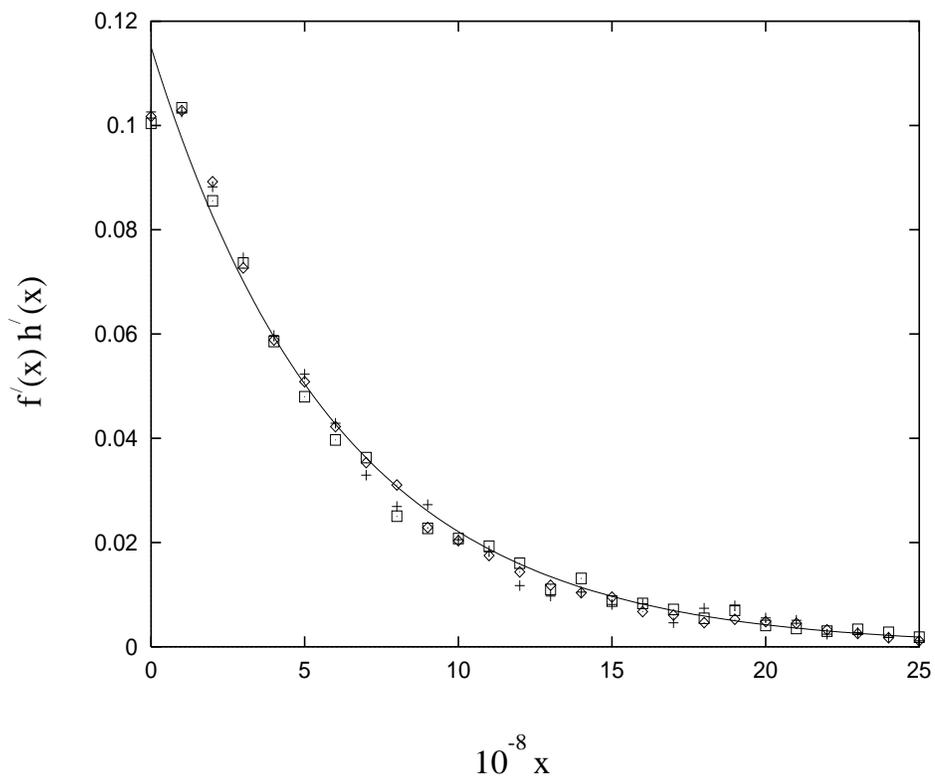
\noindent Fig 9.  The scaling function $f'(x)h'(x)$ vs $x$ for the
 rotator and mirror model on the partially occupied square lattice,
 computed at $C=C_{R_c}+0.02$ ($\diamond$), $C=C_{R_c}+0.025$
 ($+$) and $C=C_{R_c}+0.01$ ($\Box$). The curve is described by 
an exponential function, $f'(x)h'(x)=0.115 e^{-0.165 \times 10^{-8}x}$.

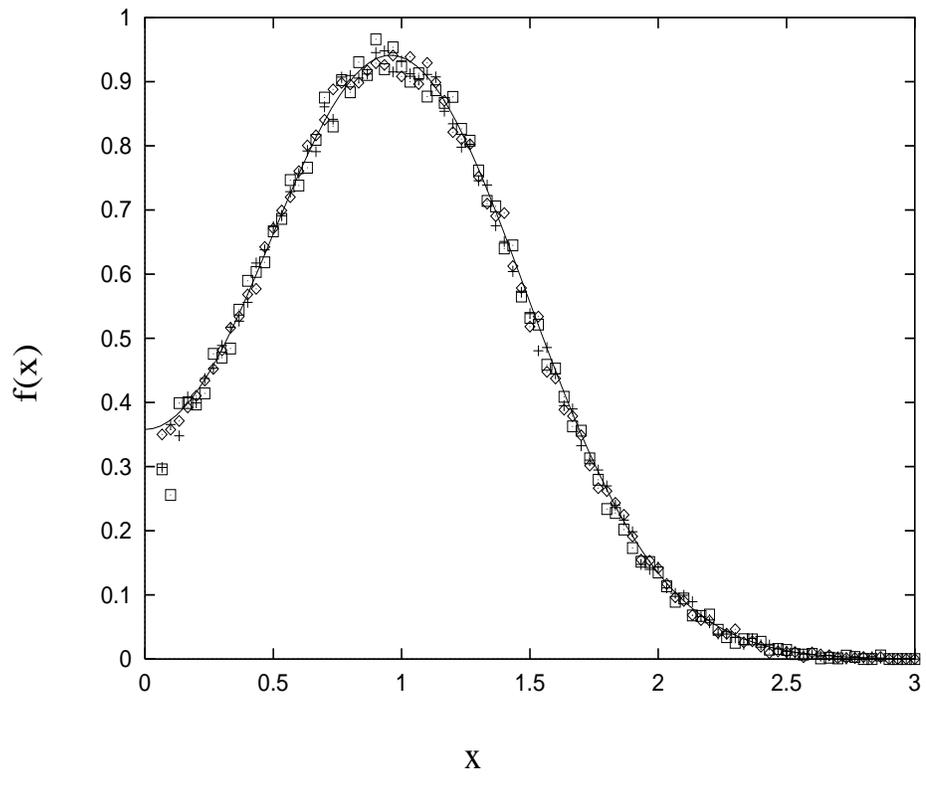
\noindent Fig 10.  The scaling function $f(x)$ vs $x$ for the 
rotator and mirror model on the fully occupied triangular lattice,
 computed at $C=C_{R_c}+0.01$ ($\diamond$), $C=C_{R_c}+0.02$ ($+
$) and $C=C_{R_c}+0.03$ ($\Box$). The curve is described by a 
double Gaussian, $f(x)=0.94 e^{-1.8(x+0.96)^2}+0.94 e^{-1.8(x-0.96
)^2}$.

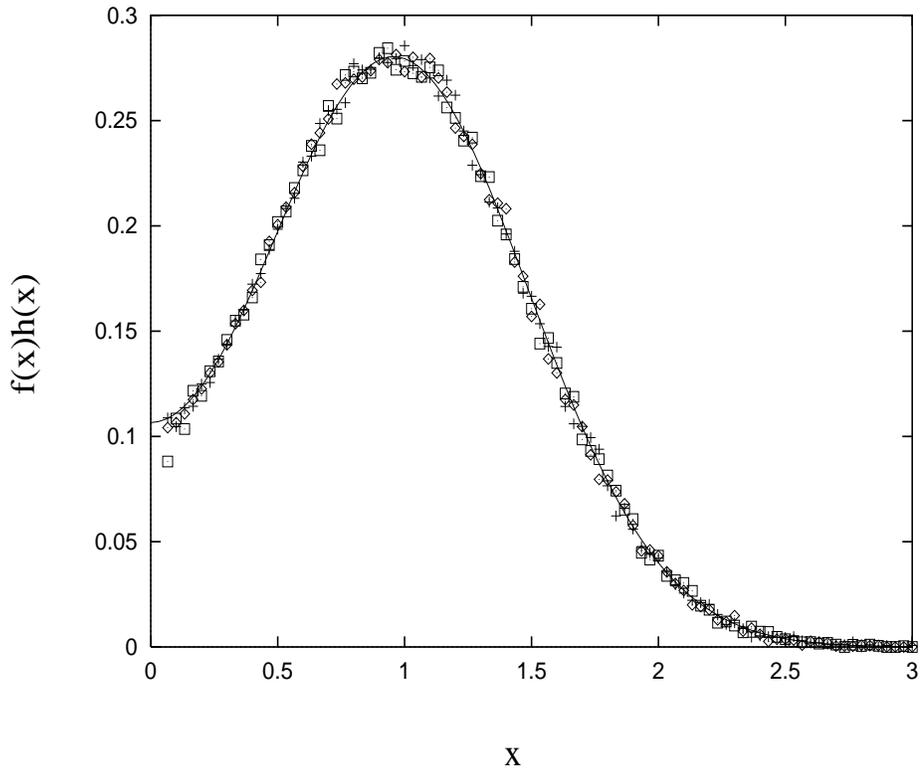
\noindent Fig 11.  The scaling function $f(x)h(x)$ vs $x$ for the
 rotator and mirror model on the fully occupied triangular lattice,
 computed at $C=C_{R_c}+0.01$ ($\diamond$), $C=C_{R_c}+0.02$
 ($+$) and $C=C_{R_c}+0.03$ ($\Box$). The curve is described by 
a double Gaussian, $f(x)h(x)=0.28 e^{-1.8(x+0.96)^2}+0.28 e^{-1.
8(x-0.96)^2}$.

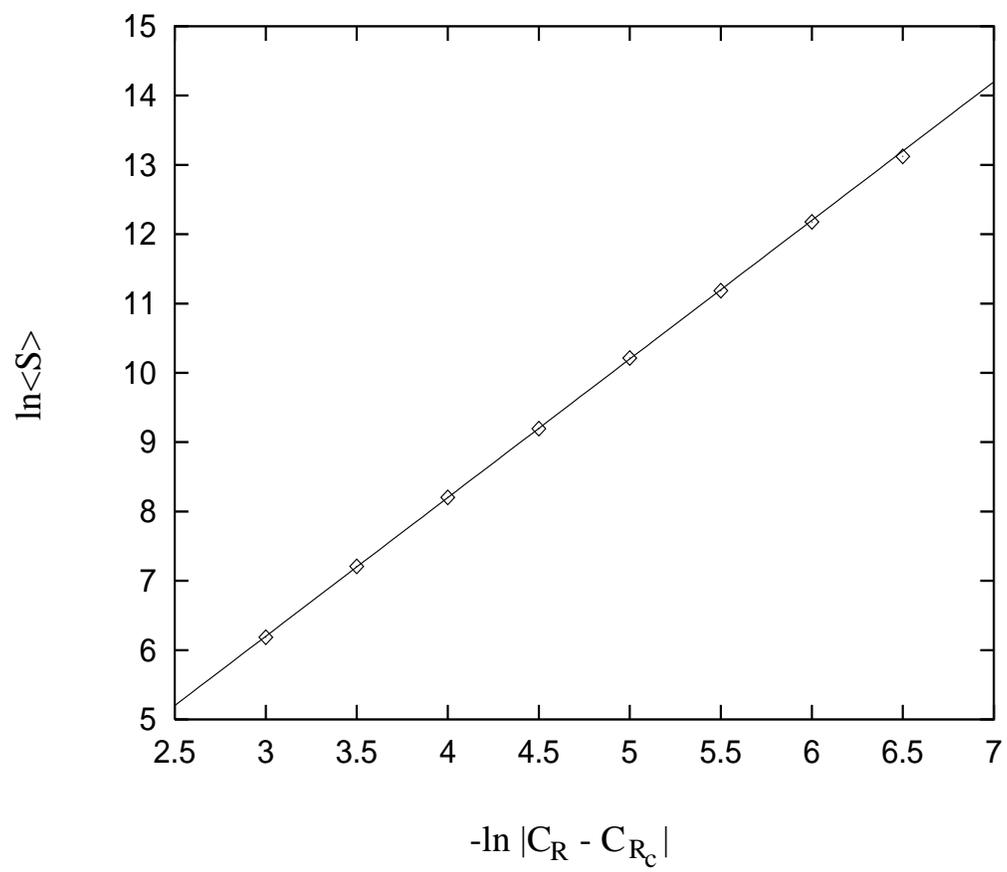
\noindent Fig. 12. $\ln<S>$ vs $-\ln|C_R-C_{R_c}|$ for the rotator
 and mirror model on the fully occupied triangular lattice. The
 slope of the fitting line is $2.00$.

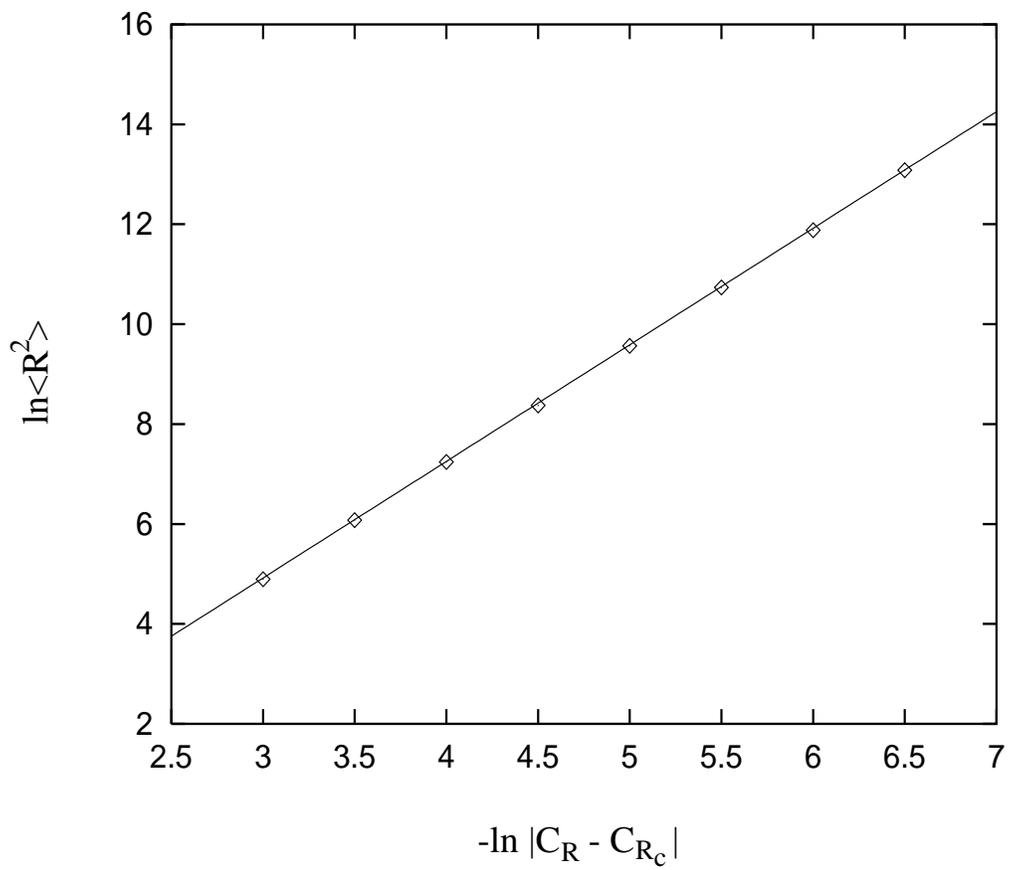
\noindent Fig. 13. $\ln<R^2>$ vs $-\ln|C_R-C_{R_c}|$ for the rotator
 and mirror model on the fully occupied triangular lattice. 
The slope of the fitting line is $2.33$.

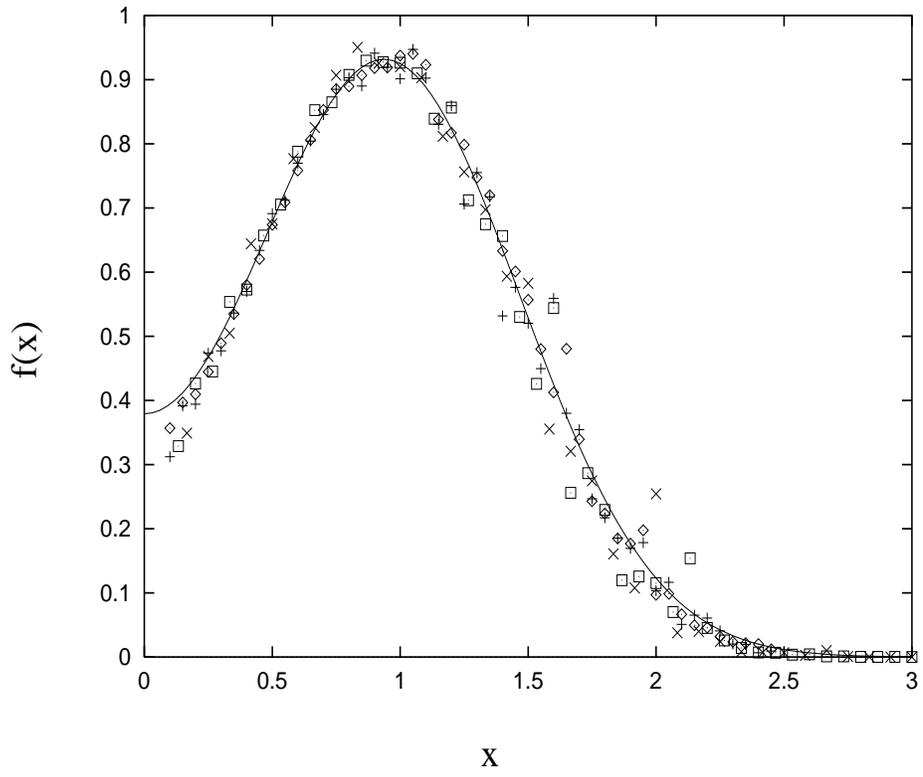
\noindent Fig. 14. The scaling function $f(x)$ vs $x$, for the 
rotator and mirorr model on the fully occupied triangular lattice
, obtained by using the histogram method, computed for $C=C_{R_c
}+0.02$ ($\diamond$), $C=C_{R_c}+0.03$ ($+$),  $C=C_{R_c}+0.04$ 
($\Box$) and $C=C_{R_c}+0.05$ ($\times$). The curve is described
 by a double Gaussian, $f(x)=0.94 e^{-2.25(x+0.96)^2} +0.94 e^{-
2.25(x-0.96)^2}$.

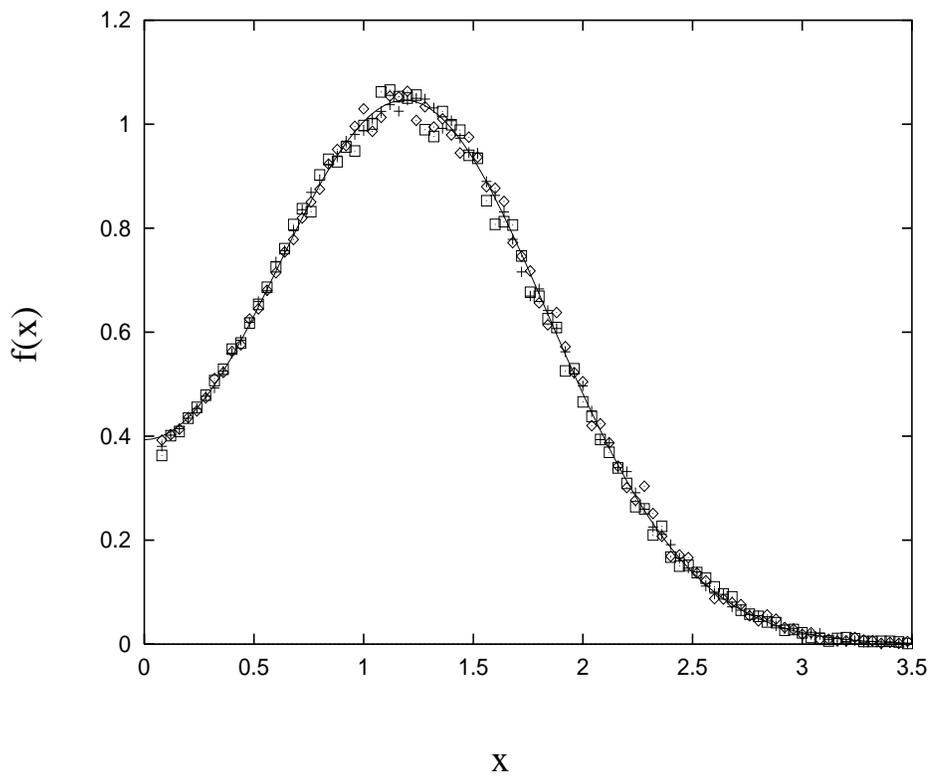
\noindent Fig 15.  The scaling function $f(x)$ v.s $x$ for the 
rotator and mirror model on the partially occupied triangular lattice,
 computed at $C=C_{R_c}+0.01$ ($\diamond$), $C=C_{R_c}+0.01
5$ ($+$) and $C=C_{R_c}+0.02$ ($\Box$). The curve is described 
by a double Gaussian $f(x)=1.05 e^{-1.18(x+1.19)^2}+1.05 e^{-1.18
(x-1.19)^2}$

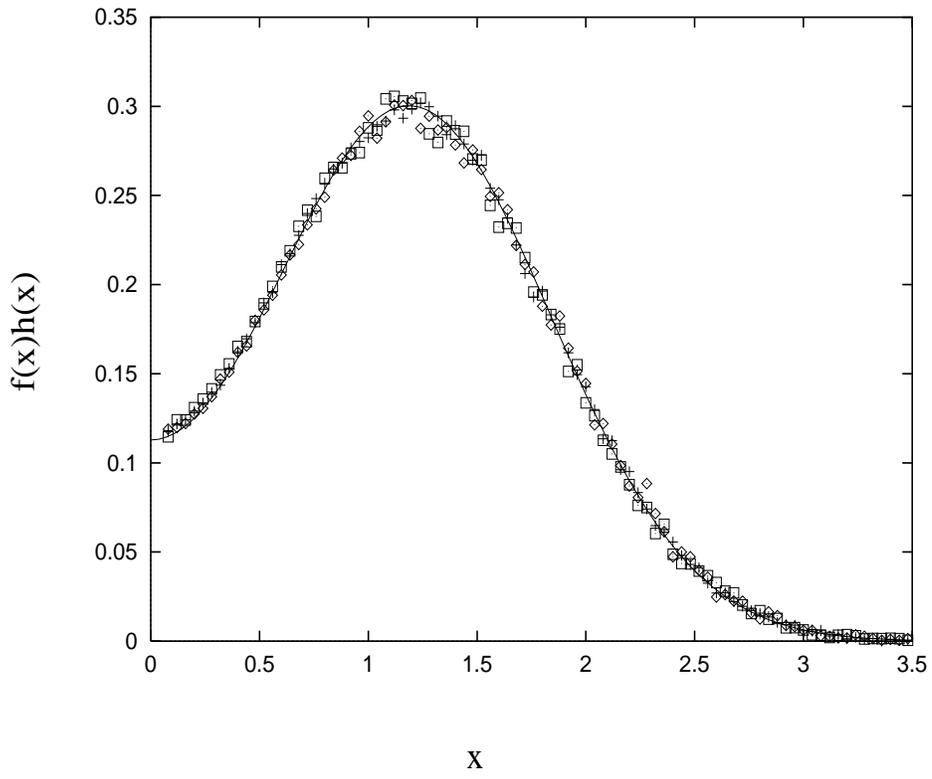
\noindent Fig 16.  The scaling function $f(x)h(x)$ vs $x$ for the
 rotator and mirror model on the partially occupied triangular 
lattice, computed at $C=C_{R_c}+0.01$ ($\diamond$), $C=C_{R_c}+0
.015$ ($+$) and $C=C_{R_c}+0.02$ ($\Box$). The curve is described
 by a double Gaussian $f(x)h(x)=0.30 e^{-1.18(x+1.19)^2}+0.30 e
^{-1.18(x-1.19)^2}$

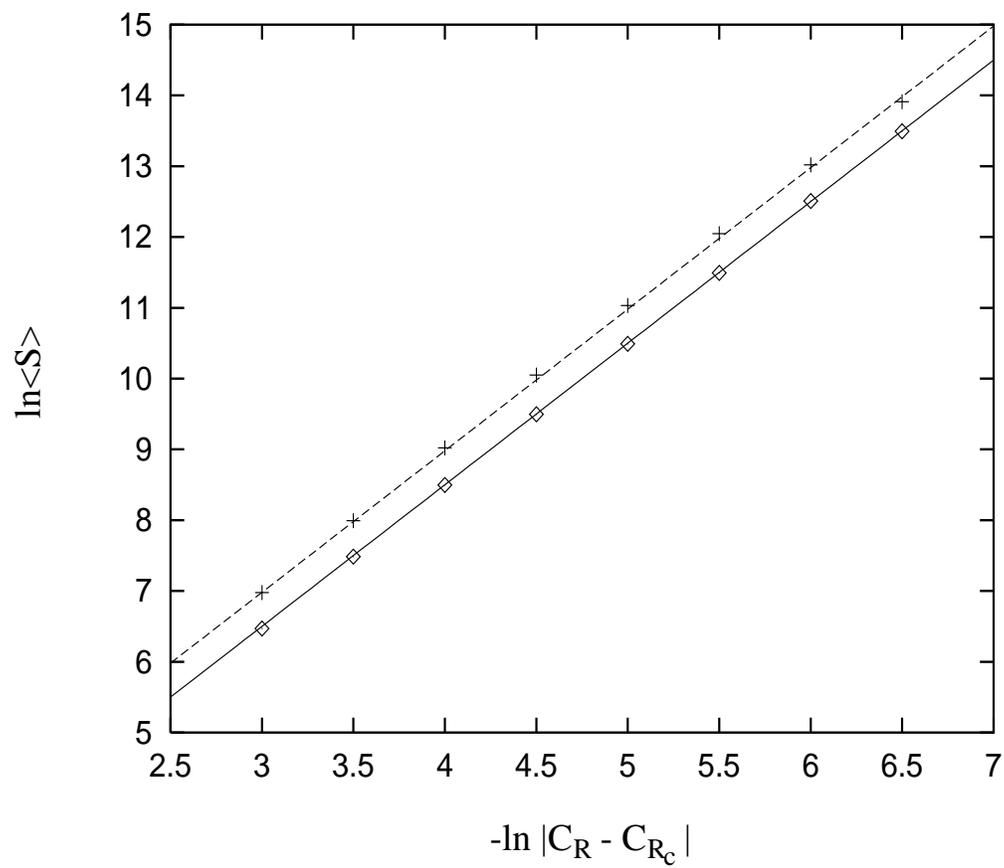
\noindent Fig. 17. $\ln<S>$ vs $-\ln|C_R-C_{R_c}|$ for the 
partially occupied triangular lattice, for $C=0.9$ ($\diamond$)and $C
=0.8$ ($+$). The slope of the fitting lines is $2.00$.

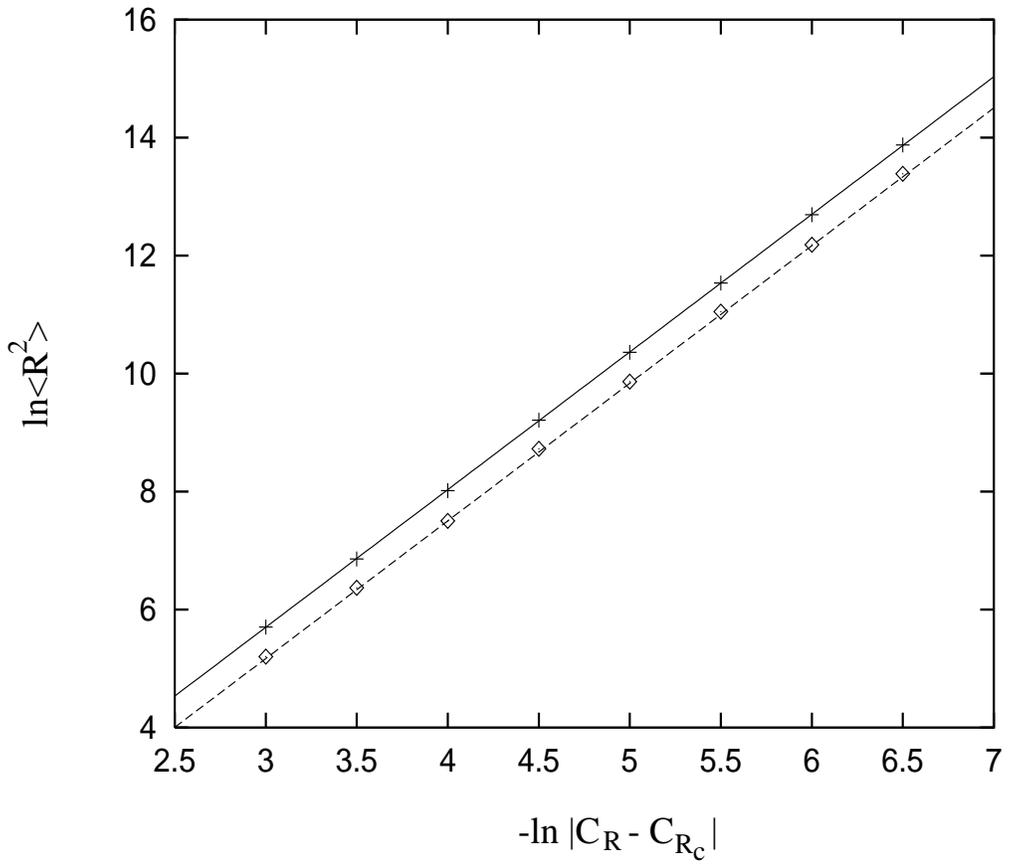
\noindent Fig. 18. $\ln<R^2>$ vs $-\ln|C_R-C_{R_c}|$ for the partially
 occupied triangular lattice, for $C=0.9$ ($\diamond$)and 
$C=0.8$ ($+$). The slope of the fitting lines is $2.33$.

\newpage
\begin{center}
{\bf {Table I:  Scaling functions and critical exponents} } 
\vskip 1.5cm

\begin{tabular}{|c|c||c|c||c|c|} \hline \hline
\multicolumn{2}{|c|}{ } & \multicolumn{4}{|c|}{Lattice} \\ \hline \hline
& & \multicolumn{2}{c|}{Square} & \multicolumn{2}{|c|}
{Triangular} \\ \hline \hline
Scaling Functions & & C = 1 & $C < 1$ & C = 1 & $C < 1$ \\ \hline \hline
& $A_1$ & 1.03 & & 0.94 & 1.05 \\ 
$f(x)$ (double Gaussian) & $\alpha_1$ & 2.25 & & 1.80 & 1.18 \\ 
& $a_1$ & 0.86 & & 0.96 & 1.19 \\ 
& & & & & (C = 0.85) \\ \hline
$h(x)$ (constant) & & 0.38 & & 0.30 & 0.28 \\ \hline \hline 
& $A'_1$ & & 0.475 & & \\ 
$f'(x)$ (exponential) & $\alpha'_1$ & & 1.65 $\cdot 10^{-8}$ & &
 \\
& $a'_1$ & & 0 & & \\ \hline
$h'(x)$ (constant) & & & 0.24 & & \\ \hline \hline
{Exponents} & & & & & \\ \hline \hline
$\tau$ & & 15/7 & 15/7 & 15/7 & 15/7 \\
$d_f$ & & 7/4 & 7/4 & 7/4 & 7/4 \\
$\sigma \; (\sigma'$) & & 3/7 & 1.6 & 3/7 & 3/7 \\
$\gamma \; (\gamma'$) & & 2 & 0.54 & 2 & 2 \\
$\rho \;(\rho')$ & & 7/3 & 0.63 & 7/3 & 7/3 \\ \hline \hline
\end{tabular}
\end{center}
\normalsize{
Scaling functions and critical exponents obtained for closed 
trajectories on
the square and triangular lattice.  $f(x) = A_1$ exp$[ - 
\alpha_1 (x - a_1)^2]$
+ $A_1$ exp[$ - \alpha_1 (x + a_1)^2]$ and $f' (x)=A'_1$ exp[ $ 
- 
\alpha'_1(x-a'_1)]$.  The critical exponents for $C = 1$ are known
 exactly,
those for $C < 1$ only numerically.  The square lattice for $C <
 1$ behaves 
exceptional if the critical point $C_{R_{c}} = C_{L_{c}} = 1/2$ 
is approached
along the line $C_R = C_L$, rather than along the line $C = 1$; 
the primed
quantities refer to this case.}

\end{document}